\documentclass[usenatbib,usegraphicx]{mn2e}
\bibpunct{(}{)}{;}{a}{}{,}

\def\gs{\mathrel{\raise0.35ex\hbox{$\scriptstyle >$}\kern-0.6em 
\lower0.40ex\hbox{{$\scriptstyle \sim$}}}}
\def\ls{\mathrel{\raise0.35ex\hbox{$\scriptstyle <$}\kern-0.6em 
\lower0.40ex\hbox{{$\scriptstyle \sim$}}}}
\defcitealias{couchw87}{CS87}
\defcitealias{zabludoffa96}{Z96}
\defcitealias{andreons02b}{Andreon \& Cuillandre's (2002)}

\title[Luminosity segregation in three clusters of galaxies]
{Luminosity segregation in three clusters of galaxies \\(A119, 2443, 2218)} 
\author[Michael.~ B.~ Pracy et al.]{
\parbox[t]{\textwidth}{
       Michael B.~Pracy$^{1,2}$, Simon P.~Driver$^2$, Roberto De~Propris$^3$, Warrick J.~Couch$^1$ and Paul E.J.~Nulsen$^4$}  
\\
\vspace*{6pt}\\
$^1$School of Physics, University of New South Wales, Sydney NSW 2052, 
Australia \\
$^2$Mount Stromlo Observatory, The Australian National University, Weston Creek, ACT 2611, Australia \\
$^3$Astrophysics Group, HH Wills Physics Laboratory, University of Bristol, Tyndall Avenue, BS8 1TL, UK\\
$^4$Harvard-Smithsonian Center for Astrophysics, 60 Garden St., Cambridge, MA 02138, USA; on leave from the  University of \\ Wollongong, NSW 2522, Australia
}

\date{Received 0000; Accepted 0000}

\pagerange{\pageref{firstpage}--\pageref{lastpage}}
\pubyear{2005} 
\begin{document}

\maketitle

\label{firstpage}
             
\begin{abstract}
We use deep wide--field $V$-band imaging obtained with the Wide Field
Camera at the prime focus of the Issac Newton Telescope to study the
spatial and luminosity distribution of galaxies in three low redshift ($0.04<z<0.2$)
clusters: Abell 119, Abell 2443 and Abell 2218.  The absolute magnitude limits
probed in these clusters are $M_{V} - 5 \log h_{0.7} = -13.3,\;
-15.4\;{\rm and} -16.7$\,mag, respectively. The galaxy population, at
all luminosities, along the line-of-sight to the clusters can be
described by the linear combination of a King profile
and a constant surface density of field galaxies. We find that, for these three clusters, the core radius
is invariant with intrinsic luminosity of the
cluster population to the above limits and thus there is no evidence 
for luminosity segregation in these clusters.
The exception is the brightest galaxies in A2218 which exhibit 
a more compact spatial distribution. We find the total projected
luminosity distribution (within $1 h_{0.7}^{-1}$\,Mpc of the cluster centre)
can be well represented by a single \citet{schechterp76} function with
moderately flat faint--end slopes: $\alpha=-1.22_{-0.06}^{+0.07}$
(A119), $\alpha=-1.11_{-0.09}^{+0.10}$ (A2443) and
$\alpha=-1.14_{-0.07}^{+0.08}$ (A2218). We perform a geometric
deprojection of the cluster galaxy population and confirm that no
`statistically significant' evidence of a change in the shape of the
luminosity distribution with cluster-centric radius exists. Again, the 
exception being A2218 which exhibits a core region with a flatter faint--end slope.
\end{abstract}


\section{Introduction}
Galaxies of different types in clusters are known to have different
projected spatial distributions. This was realized by
\citet{oemlera74}, who showed that less luminous galaxies have a more
extended profile than the more massive ellipticals. \citet{melnickj77}
and \citet{dresslera80} identified what is now known as the
`morphology-density' relation, where the relative fractions of
elliptical, lenticular (S0) and spiral galaxies depend on the surface
density, while \citet{whitmoreb93} argued that these trends are better
correlated with cluster-centric radius. In a recent comprehensive
study of an ensemble cluster built from 59 nearby rich clusters
\cite{bivianoa02}, demonstrated clear segregation between ellipticals,
early and late-type spirals. This is also seen in a single HST mosaic
of Abell 868 by \cite{drivers03} which concludes that cluster cores are
devoid of, or at least depleted, in late-type systems.

However, as well as morphological segregation, evidence is also
emerging for luminosity segregation. \citet{roodh68} argued that
dwarfs were less concentrated than giants in the Coma cluster;
\citet{capelatoh81} detected a mass-density relation in Abell 196;
\citet{yepesg91} studied luminosity segregation in a number of
clusters and found that the degree of segregation correlates with the
dynamical state of the cluster. The study of \cite{fergusonh89} in
Virgo and Fornax demonstrated that dwarf ellipticals were highly
concentrated leading to a division of the dwarf population into
distinct strongly clustered nucleated dwarf ellipticals and a
distributed population consisting of non-nucleated dwarf ellipticals
and dwarf irregulars. In the Coma cluster, \citet{loboc97} and
\citet{kashikawan98} found evidence for strong luminosity segregation,
with the giants being clumped in two substructures while the dwarfs
traced a more diffuse and regular distribution. \citet{andreons02a}
argued that some form of mass segregation is also at work in the
$z=0.31$ cluster AC 118 (also known as Abell 2744). 
The giant ellipticals and lenticulars may also
be kinematically segregated \citep{steinp97}, suggesting that these
objects are the original kernel of the clusters while spirals and
dwarfs are comparatively late arrivals. Conversely, \citet{bivianoa02}
find that the only evidence for luminosity segregation is for
ellipticals outside of substructures in their ensemble cluster.

\citet{smithr97} and \citet{drivers98} found that the dwarf-to-giant
ratio shows a trend with density and, because of the approximately
spherical shape of clusters, with radius. This lead
directly to the idea of a dwarf-density relation \citep{phillippss98}.
Together with previous work \citep{loboc97,kashikawan98} this may
suggest that dwarfs are especially affected by the cluster
environment, as one would expect for such fragile objects.

\citet{pracym04} have recently investigated luminosity segregation
using a wide, deep mosaic of HST images of Abell 2218
and found evidence that dwarf galaxies avoid the central regions of this
cluster and trace a more spatially extended distribution. A similar
result was found for the NGC 5044 group by \citet{mathewsw04} and
for a sample of loose groups by \citet{girardim03}. If the segregation
for dwarfs is real, it may originate from initial conditions, where low
luminosity galaxies are only now in-falling into clusters (e.g., Croton et
al. 2005), or it may be due to processes internal to clusters, such as
tidal disruption and galaxy harassment. For these reasons, it is important
to investigate the existence of luminosity segregation for dwarfs in a
broader range of objects and to study its correlation with cluster
properties. This is now feasible by wide--field imaging of relatively
nearby clusters with panoramic mosaic cameras on 2m telescopes, and we
present here the results of such a study for three clusters observed from
the Isaac Newton 2.5m Telescope with the Wide Field Camera.

In this paper we use relatively--deep wide--field imaging
of three galaxy clusters in the redshift range $0.04<z<0.2$ to 
measure the galaxy surface density in intervals of
luminosity, recover the overall luminosity distributions and the
deprojected luminosity distributions to explore the spatial
segregation of the galaxy population with luminosity.
The absolute magnitude limits probed in these clusters are 
$M_{V} - 5 \log h_{0.7} = -13.3,\; -15.4\;{\rm and} -16.7$\,mag.
In Section 2 we introduce the data and describe its reduction and the detection
analysis and classification strategy. In section 3 we fit a King
profile plus constant offset in intervals of luminosity to
simultaneously determine both the cluster population profile and the
non-cluster foreground/background level. In Section 4 we derive the
projected luminosity distributions in these clusters and in Section
5 we perform a geometric deprojection to recover the `true' luminosity 
distribution of the galaxy population. We summarise our
findings in Section 6. Throughout we adopt a ${\Omega}_{M}=0.3$,
${\Omega}_{\Lambda} =0.7$ and $H_{0}=70$\,km\,s$^{-1}$\,Mpc$^{-1}$
cosmology.
\section{The data}

\subsection{The Observations}
The sample consists of three Abell clusters: A119 (Richness=1,
BM=II-III, z=0.044), A2443 (Richness=2, BM=II, z=0.108) and A2218
(Richness=4, BM=II, z=0.181). The observations were obtained on the
nights of $2^{nd}$ \& $3^{rd}$ September 2000 using the Wide Field Camera
(WFC) mounted at the prime focus of the Isaac Newton Telescope (INT).
The WFC consists of a mosaic of four $2048 \times 4096$ thinned EEV
CCDs with a plate scale of 0.333 arcsec/pixel. The total sky coverage
is 0.287\,deg$^2$ per pointing.  The imaging of each cluster
consists of four partially overlapping pointings with the WFC, the
exception being the higher redshift cluster A2218 which is a mosaic
of just 2 pointings.  Each cluster was imaged through the $V$ filter
with an exposure time of 1200s. A summary of the observations is given
in Table \ref{tab:summary}.
\begin{table*}
\setcounter{table}{0}
\caption{\label{tab:summary}Data characteristics.}
\begin{tabular}{|c|c|c|c|c|c|c|c|c|c|c|c|c|c|} \hline
Cluster  &   RA &  DEC   &  Redshift & Exp & Seeing & No. Fields &  Area  & Area &  BCG$\dagger$ \\
         &  \multicolumn{2}{c}{(J2000.0)} &  & (sec) & ($''$) & & $(\deg^2)$ & ($h_{0.7}^{-2}$\,Mpc$^{2}$) & (mag) \\ \hline
Abell 119    & $00^{\rm h}56^{\rm m}21^{\rm s}$  & $-01^{o}15^{'}47^{''}$ & 0.044   &  1200         & 1.14  & 4  &   0.85 & 9.7  &   13.73     \\ 
Abell 2443    & $22^{\rm h}26^{\rm m}07^{\rm s}$  & $+17^{o}20^{'}17^{''}$ & 0.108   &  1200         & 1.18 & 4  &   0.82   & 50.4 & 15.07     \\ 
Abell 2218    & $16^{\rm h}35^{\rm m}54^{\rm s}$  & $+66^{o}12^{'}00^{''}$ & 0.181   &  1200         & 0.94  & 2  &   0.45   & 120.5 & 16.40    \\ \hline
\end{tabular}

\noindent $\dagger$ Brightest Cluster Galaxy.
\end{table*}

\subsection{Data reduction}
The data reduction was performed by the Cambridge Astronomical Survey
Unit (CASU) and full details of the pipeline procedure can be found in
\citet{irwinm01}. In summary, the data are first bias subtracted and
trimmed. Bad pixels and columns are interpolated over using data from
neighbouring regions. All four chips are then corrected for non-linear
behaviour in the two analogue-to-digital converters.  The data are
then flat fielded using master sky flats and a gain correction is 
applied so that all the CCDs have the same zero-point. 
Finally, an astrometric solution is derived by
matching bright stars in the field-of-view of each chip to the Guide
Star Catalog.

\subsection{Photometric calibration}
Four \citet{landolta92} standard star fields (SA92,
SA95, SA110 and SA113) were observed at various air-masses throughout
each night.  For each observation of a standard star a zero-point was
calculated:
\begin{equation}
\label{eqn:zero}
{\rm ZP_{star}}=m+2.5\log{f \over t}
\end{equation}
where $m$ is the magnitude of the standard star given by
\citet{landolta92}, $f$ is the star counts in ADUs as measured
via {\sc SExtractor} fixed aperture magnitude ($7\arcsec$ radius) and $t$
is the exposure time (10s).  These zero-points were fitted with a
double-linear function in airmass and $(B-V)$ colour (taken from
\citealt{landolta92}) to derive the extinction coefficient $(a_{\rm
air})$ and an above-atmosphere zero-point $a_{0}$.
\begin{equation}
\label{eqn:zeroabove}
{\rm ZP_{star}}=a_{0}+a_{\rm air}(\sec Z)+a_{\rm col}(B-V)
\end{equation}
We then assigned to each field a final zero-point $ZP=a_{0}+a_{air}
{\sec Z}$.  Fig.~\ref{fig:standard} shows the data used for the $V$-band calibration
on the night of $3^{\rm rd}$ September 2000, the top panel shows the
colour-corrected zero-point versus airmass and the bottom panel shows
the airmass-corrected zero-points versus $(B-V)$ colour.  The fit
(equation \ref{eqn:zeroabove}) to the data is represented by the {\it solid line} and
the RMS of the data (top panel) is $\pm 0.06$\,mag.
\setcounter{figure}{0}
\begin{figure}
\includegraphics[height=8cm,angle=-90]{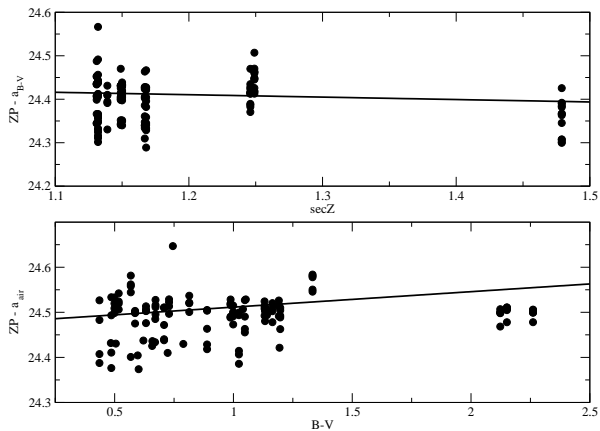}
\caption{\label{fig:standard}$V$-band standard star calibration for $3^{\rm rd}$ September
2000. Colour-corrected zero-points versus airmass (top) and
airmass-corrected zero-points versus colour (bottom)}
\end{figure}

\subsection{Object detection and photometry}
Objects were detected automatically using the {\sc SExtractor }
package \citep{bertine96}.  {\sc SExtractor } detects objects as
groups of connected pixels which exceed a certain threshold.  A
criterion of 8 connected pixels above an isophotal threshold of $\mu_{0}=26.0$\,mag\,arcsec$^{-2}$ was used.  The {\sc best\_mag} parameter
was used to measure total galaxy magnitude (hereafter denoted $V$);
this corresponds to a \citet{kronr80} extraction aperture, except for
crowed regions where an extrapolated isophotal magnitude is used. The
Kron extraction aperture is set to 2.5 times $R_{K}$ where $R_{K}$ is
the first moment of the image distribution. In regions where pointings 
overlap, the duplicate objects were removed from the catalogs.

\subsection{Exclusion regions}
After the initial object detection each field was visually inspected
and any CCD defects, very bright stars, diffraction spikes and
satellite trails causing spurious detections were identified.  Circular and
rectangular regions enclosing these areas were defined and excluded
from the catalogs.  In addition, the areas within 20 pixels of the CCD
edge and the vignetted corner of CCD3 were also excluded.

\subsection{Object classification}
We used the position of objects in the central surface brightness--magnitude ($\mu_o$--$V$) plane to classify objects as either
galaxies, stars or cosmic rays. The distribution of a subset of detected
objects in this plane, for each cluster, is illustrated in Fig.~\ref{fig:muv}. The
central surface brightness was calculated in a circular aperture with
an area equal to that of the detection criterion (i.e., 8 pixels).
\setcounter{figure}{1}
\begin{figure}
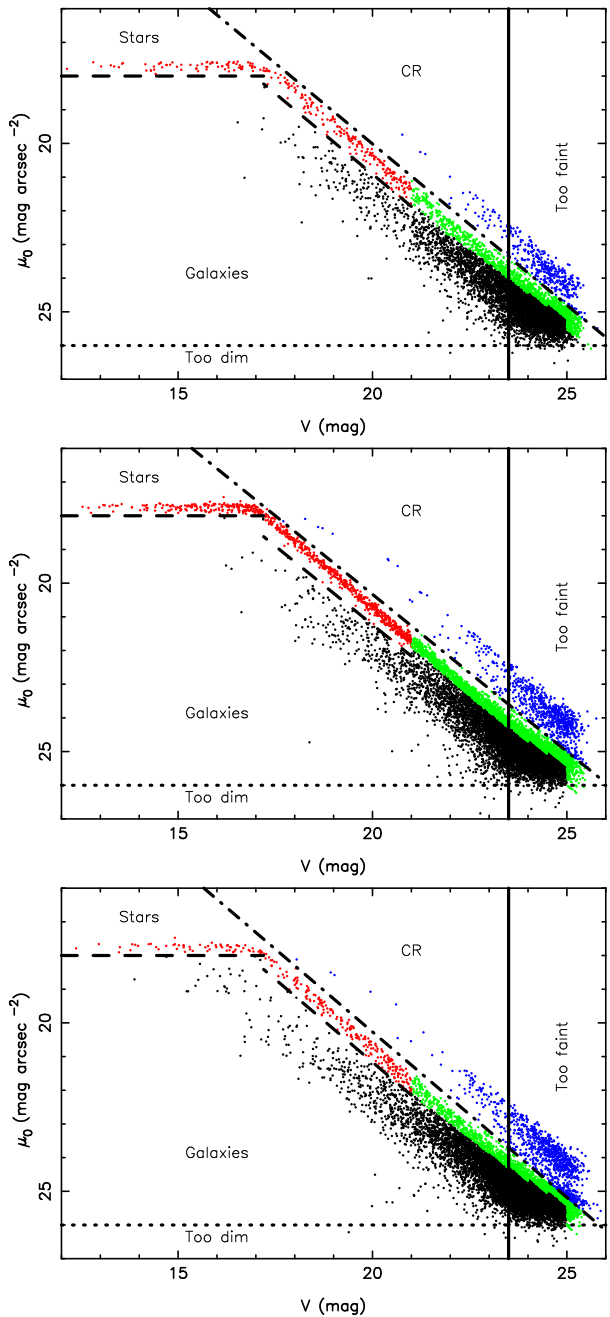

  \begin{center}
    \begin{minipage}{0.47\textwidth}
          \includegraphics[height=8.0cm, angle=-90, trim=0 0 0 0]{figure2a.ps}
          \includegraphics[height=8.0cm, angle=-90, trim=0 0 0 0]{figure2b.ps}
	  \includegraphics[height=8.0cm, angle=-90, trim=0 0 0 0]{figure2c.ps}
     \end{minipage}
    \caption{\label{fig:muv}Distribution of detected objects in the
central surface brightness--magnitude plane.  {\it Blue points} are
objects classified as cosmic rays, {\it red points} are objects
classified as stars directly from their positions in this plane, {\it
green points} are objects classified as stars by extrapolation from
the brighter star counts and {\it black points} are the objects
classified as galaxies. Top panel: A2218. Middle panel: A2443. Lower
panel: A119. Only one in six objects, randomly selected, are displayed.}
    \end{center}
\end{figure}  

\subsubsection{Cosmic ray rejection}
In Fig.~\ref{fig:muv} a group of objects with a high central surface brightness
(at a given magnitude) is clearly discernible ({\it upper
right}). These objects, which have surface brightnesses higher than
that of stars, are cosmic rays. We therefore define a region in the
$\mu_o$--$V$ plane such that:
\begin{equation}
\label{eq:cosmicray} 
\mu_o\le aV+b
\end{equation} 
and classify all objects in this region as cosmic rays.  Equation (\ref{eq:cosmicray})
 is shown as the {\it dot-dashed line} in Fig.~\ref{fig:muv}. The slope ($a$)
and intercept ($b$), were chosen separately for each cluster, to best
match the data. The objects classified as cosmic rays are shown as
{\it blue points} in Fig.~\ref{fig:muv}.
\subsubsection{Star-galaxy separation}
The process of star-galaxy separation begins by identifying saturated stars in the catalogs.  Stars
brighter than $V\approx 17.2$\,mag are saturated, these are clearly
identifiable in the $\mu_o$--$V$ plane as a horizontal locus
of points with $\mu_o<18.0$\,mag\,arcsec$^{-2}$. We classify
objects with:
\begin{equation}
\label{eq:flooded}
\mu_o<18.0 {\;\rm and\;} V\le 17.2
\end{equation}
as flooded stars ({\it horizontal dashed line } in Fig.~\ref{fig:muv}). The
stellar locus can be seen in Fig.~\ref{fig:muv} as a diagonal locus of points
({\it red}) with a higher surface brightness than the overall galaxy
population ({\it black points}), and extending from $V\approx 17.2$ to
$V\approx 21$\,mag.  We therefore define a line in the
$\mu_o$--$V$ plane:
\begin{equation}
\label{eq:stargal}
\mu_o = aV+b' {\;\rm and\;} V>17.2 {\;\rm and\;} V\le 21.0
\end{equation}
to pass between these populations, and we use it as a divider to
separate stars and galaxies (see {\it diagonal dashed line} in Fig.~\ref{fig:muv}).
All objects with $V\le 21$\,mag which are classified as stars are shown
in {\it red} in Fig.~\ref{fig:muv}.

For objects fainter than $V \approx 21$\,mag star-galaxy separation
becomes problematic. At these magnitudes the stellar locus merges
with that of the overall galaxy population and the two can no longer
be distinguished. In order to perform star-galaxy classification
faintward of $V=21$\,mag we use a similar method to that outlined in
\citet{liskej03}.  Since the (logarithmic) slope of the star counts
should remain roughly constant to $V\approx 24$\,mag
\citep{kummelm01}, we can use the star counts derived from the bright
objects in the catalog and extrapolate them to derive the expected number
of star counts at fainter magnitudes. We then classified objects as
stars based on their position in the $\mu_o$--$V$ plane
(objects with the lowest value of $\mu_o-mV$) until we had
obtained the predicted number of stars. Although this method will
result in some individual objects having incorrect classifications,
the overall statistical proprieties of the catalogs will be
correct. The objects that have been classified as stars in this way
are shown in {\it green} in Fig.~\ref{fig:muv}.

\subsubsection{Completeness}
At this point every object in the catalogs has been classified as
either a star, galaxy or cosmic ray. We are limited to galaxies 
with a central surface brightness (over an area of 8 pixels) of
$\mu_o\le 26$\,mag\,arcsec$^{-2}$ and from Fig.~\ref{fig:muv} we see
that objects with central surface brightnesses close to this limit
only occur in significant numbers at $V > 23.5$\,mag, indicating the
beginning of detection incompleteness \citep{garillib99}. We therefore, 
define an apparent magnitude limit of $V=23.5$\,mag.


\section{Radial Profiles}
The wide field-of-view provided by the WFC mosaics enable us to survey
the radial distribution of galaxies beyond the domain of the cluster, 
well out into the surrounding field. To explore this, we plot for each
cluster the galaxy-surface density against cluster-centric
radius. This is achieved by deriving the galaxy counts in concentric
annuli centred on the brightest cluster galaxy. The surface density
was calculated taking into account the area of each annulus which
encompasses the unmasked field-of-view of the available CCD area (i.e.,
corrected for any exclusion regions which intersect the annulus). We
also corrected for the `diminishing area effect' \citep{drivers98b}
whereby the area over which faint objects can be detected is reduced
by the presence of brighter objects. To calculate this effect we used
the {\sc SExtractor } parameter {\sc ISOAREA} -- which returns the
total area assigned to an object by {\sc SExtractor} -- to calculate
the amount of area occupied by brighter objects.

The {\sc SExtractor best} magnitudes were corrected for galactic
extinction using the maps of \citet{schlegeld98}.  The observed radial
galaxy surface density profiles for A2218, A2443 and A119 are shown
in Fig.~\ref{fig:rad1}, Fig.~\ref{fig:rad2} and Fig.~\ref{fig:rad3}, respectively, as {\it open squares}. The
error bars on the points are those expected from purely Poisson
statistics.

\setcounter{figure}{2}
\begin{figure*}
\includegraphics[height=15cm,angle=-90]{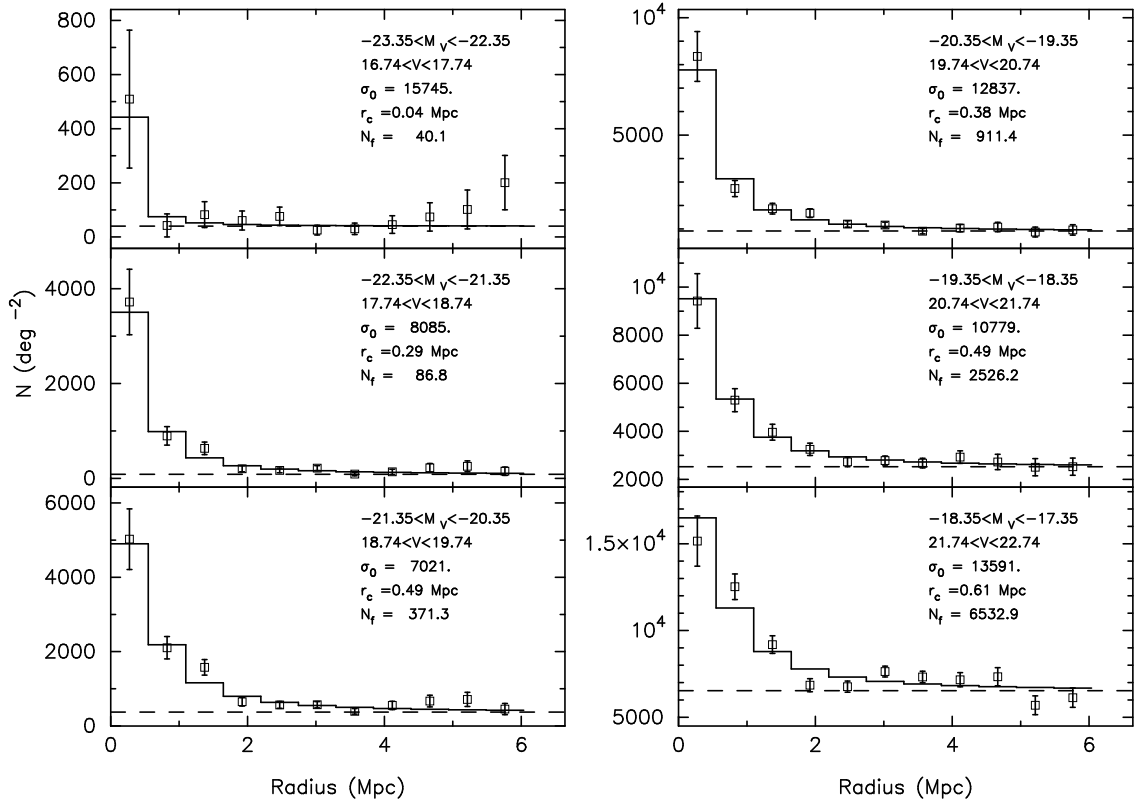}
\caption{\label{fig:rad1}The surface density of galaxies in A2218 as a function of
cluster-centric radius for a series of magnitude intervals.  The {\it
solid curves} are the best fitting `King + constant' profile given by
equation (\ref{eq:kingint}). The (apparent) magnitude intervals and their
corresponding absolute magnitude intervals (for the cluster redshift)
as well as the best fitting values of the fit parameters are tabulated
in each panel. The {\it dashed lines} show the value of the $N_{f}$ parameter 
from equation (\ref{eq:kingint}).}
\end{figure*}
\setcounter{figure}{3}
\begin{figure*}
\includegraphics[height=15cm,angle=-90]{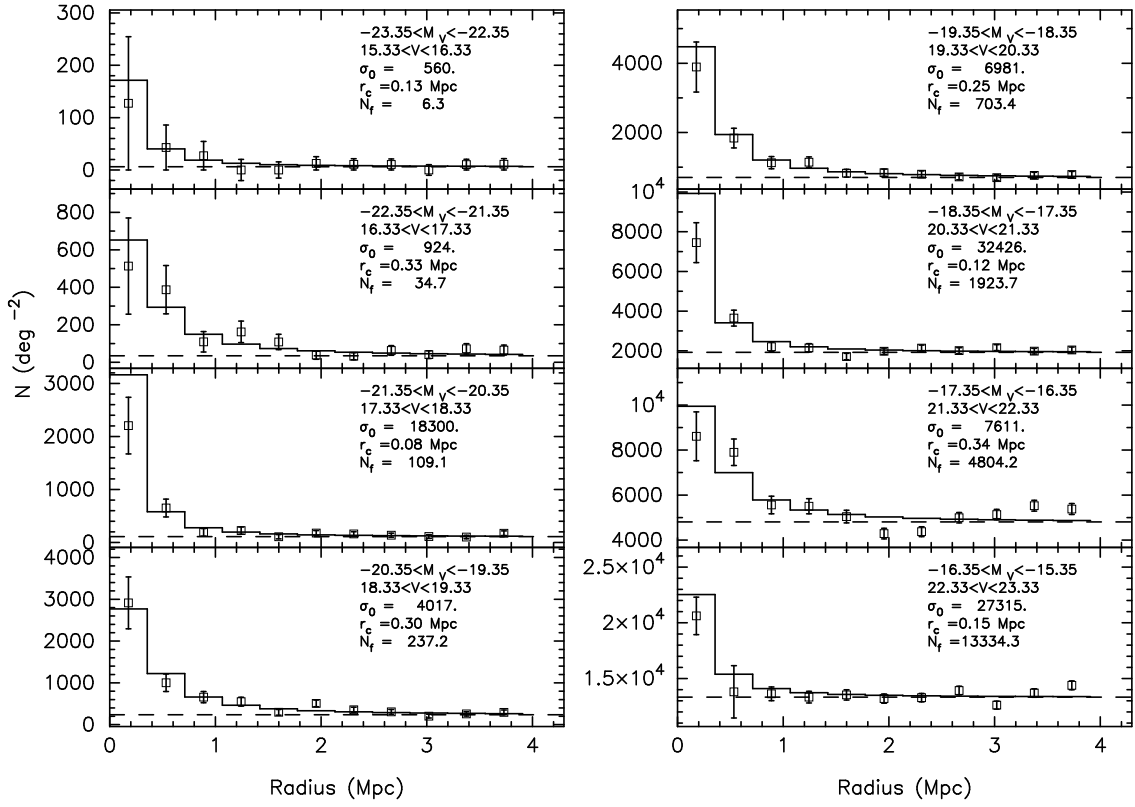}
\caption{\label{fig:rad2}Same as Fig.~\ref{fig:rad1} except for A2443} 
\end{figure*}
\setcounter{figure}{4}
\begin{figure*}
\includegraphics[height=15cm,angle=-90]{figure5.ps}
\caption{\label{fig:rad3}Same as Fig.~\ref{fig:rad1} except for A119} 
\end{figure*}

\subsection{Cluster profiles}
The radial profiles in Figs.~\ref{fig:rad1}--\ref{fig:rad3} represent the superposition of the
cluster surface density profiles and a constant `field' galaxy surface
density. We elect to describe the cluster surface density in functional form
by a \citet{kingi62} profile, plus a constant surface density of `field' galaxies,
thus:
\begin{equation}
\label{eq:king}
\sigma(r) = {\sigma_{0} \over 1+\bigl({r \over r_{c}}\bigr)^2} + N_f
\end{equation}
where $\sigma(r)$ represents the radially ($r$) dependent 
number-counts along the line-of-sight. The first term in
equation (\ref{eq:king}) represents the projected distribution of
cluster galaxies and the second term the superimposed `field'
population. 
In Figs.~\ref{fig:rad1}--\ref{fig:rad3} we have binned the galaxy number counts in radial annuli. 
We therefore need to integrate our fitting function (equation \ref{eq:king}) over the
annuli to obtain the average surface density, which gives
\begin{eqnarray}
{1 \over \pi(r_{max}^2-r_{min}^2)}\int_{r_{min}}^{r_{max}}2\pi r \sigma(r)dr= \nonumber \\
\label{eq:kingint}
\left. {1 \over r_{\rm max}^2 - r_{\rm min}^2} \left[ N_f r^2 + \sigma_0 r_c^2 \log{(1 + (r / r_c)^2)} \right]\right|_{r_{\rm min}}^{r_{\rm max}}
\end{eqnarray}
where $r_{min}$ and $r_{max}$ are the inner and outer radial boundaries of the bin, respectively. The fitted profiles to the measured galaxy
surface density profiles are shown as the {\it solid lines} in
Fig.~\ref{fig:rad1}--\ref{fig:rad3} in 1\,mag intervals of $V$.  Overall, the surface density of galaxies are well described by equation (\ref{eq:king}).

In Fig.~\ref{fig:core} we show the King profile core radii [given by the parameter
$r_c$ in equation (\ref{eq:kingint})] as a function of absolute $V$-band magnitude, the 
errors on the points are calculated directly from the covariance matrix returned
from $\chi^2$ minimisation. In the case of the clusters 
A2443 ({\it middle panel}) and A119 ({\it lower panel}) 
we find that the core radius is essentially independent
of magnitude, with both clusters having core radii of $r_c \approx$
0.2--0.5\,Mpc at all luminosities. The best fitting slopes are given
by $-0.001\pm 0.033$\,Mpc\,mag$^{-1}$ and $0.058\pm 0.040$\,Mpc\,mag$^{-1}$, respectively -- both consistent with zero at $\sim 1\sigma$. 
In A2218, we find that the brightest galaxies have a smaller core radius then their fainter counterparts, with
marginal evidence that this trend -- increasing core radius with decreasing luminosity -- continues for the intermediate population.
The best fitting slope is given by $0.086\pm 0.030$\,Mpc\,mag$^{-1}$. However, if the brightest
points ($M_{V}< -21$\,mag) are removed, the slope becomes 
$0.051\pm 0.054$\,Mpc\,mag$^{-1}$ which is consistent with
zero at less than the $1\sigma$ level. This trend is generally consistent 
with  \citet{pracym04} who found
that the spatial distribution of galaxies in A2218 is more extended for the
lower luminosity populations. Unfortunately the data do not extend to their
dwarf and ultra-dwarf regimes.

\setcounter{figure}{5}
\begin{figure}
\includegraphics[height=8cm,angle=-90]{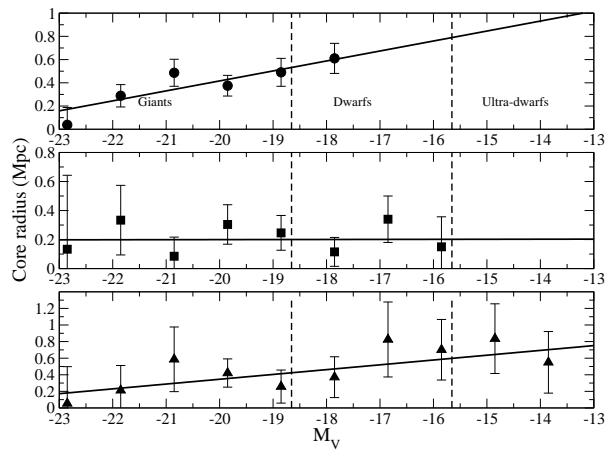}
\caption{\label{fig:core}Core radii [$r_{c}$ from equation (\ref{eq:kingint})] versus absolute $V$-band magnitude. Top panel: A2218. Middle panel:
A2443. Lower panel: A119. The best fitting slopes are displayed as {\it solid lines}.  The `Giant', `Dwarf' and
`Ultra-dwarf' regimes, as defined by \citet{pracym04}, are delineated by the {\it dashed lines}. These 
definitions have been adjusted due to differences in the filters and the assumed cosmology.} 
\end{figure}

\subsection{Reference field counts}
In order to study the cluster galaxy population we need to remove the
contribution to the counts along the line-of-sight from
foreground and background `field' galaxies. To do this we use the
standard technique of statistical field subtraction. However rather
than use data obtained from off-cluster pointings we are now able to
use the counts derived from the radial fits ({\it i.e.,} $N_f$ in
equation \ref{eq:kingint}).

The field galaxy number counts derived in this way are shown in the
{\it top panel} of Fig.~\ref{fig:backcounts}, where we have plotted the counts from the
A2218 field ({\it filled squares}), the A2443 field ({\it filled
circles}) and the A119 field ({\it stars}), separately. For comparison
we also plot the galaxy number counts from the Millennium Galaxy
Catalog (MGC) of \citet{liskej03}, which we have converted from the
$B$-band to the $V$-band using the mean field galaxy colour $(B-V)
= 0.94$ \citep{norbergp02}. We note that this colour is
only strictly valid for $V < 18$\,mag; however, the number counts agree
quite well over the entire range of luminosities.  The `field' counts
derived for the low redshift cluster A119 generally have larger error
bars and much larger scatter than those derived for the other two
clusters. This is expected since the radial
profiles for this cluster extend to a radius of only $\sim 2$\,Mpc. 
The scatter in the points can be seen more clearly in the {\it lower panel} 
of Fig.~\ref{fig:backcounts} where we show the galaxy counts relative to a
$N\propto 10^{0.45V}$ relation.

\setcounter{figure}{6}
\begin{figure}
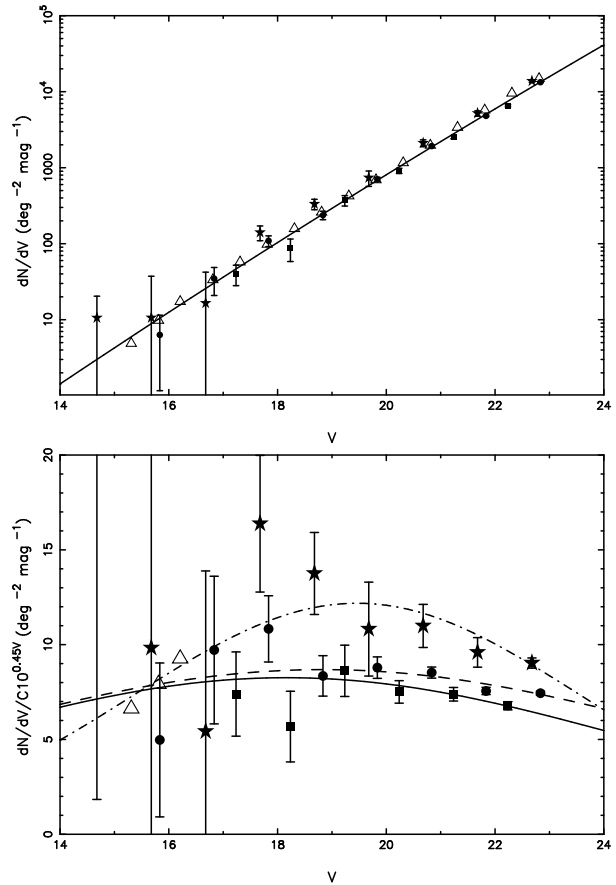

  \begin{center}
    \begin{minipage}{0.47\textwidth}
          \includegraphics[height=8.0cm, angle=-90, trim=0 0 0 0]{figure7a.ps}
          \includegraphics[height=8.0cm, angle=-90, trim=0 0 0 0]{figure7b.ps}
     \end{minipage}
    \caption{\label{fig:backcounts}Top: `field' galaxy counts derived from equation (\ref{eq:kingint}, see
text for details); {\it filled squares} are the reference `field'
counts from the A2218 field; {\it filled circles} for the A2443 field
and {\it stars} for the A119 field. The {\it open triangles} are the
`field' galaxy counts from the MGC. The {\it solid line} is a quadratic fit
to the counts from all three clusters. Bottom: `field' galaxy counts
shown relative to a $10^{0.45V}$ relation. The quadratic fits to the
counts are shown as the {\it solid line} (A2218), {\it dashed line}
(A2443) and {\it dot-dash line} (A119).  }
    \end{center}
\end{figure}  

In order to give a smooth representation of the `field' counts we
fitted a quadratic function to them. Since `field' galaxy counts can vary significantly
between fields, a quadratic fit was performed separately for each set
of reference `field' counts. These fits are shown in the lower panel
of Fig.~\ref{fig:backcounts} as the {\it solid, dashed} and {\it dot-dashed} lines for
A2218, A2443 and A119, respectively.  For illustrative purposes a simultaneous fit
to all three data sets is shown as the {\it solid line
} in the top panel of Fig.~\ref{fig:backcounts}.  We also used the MGC counts at $V < 18$\,mag 
to provide a bright end `anchor' to the fit -- we artificially
introduced errors of 10\% on these points so as not to allow the MGC
points to influence the fit at the {\it critical} faint end. The quadratic fits yield:
\begin{eqnarray}
\label{eq:backcounts}
& {dN\over dV} & = 10^{a+bV+cV^2} \\ 
&a& = -7.83486, -7.64871, -10.87945  \nonumber \\ 
&b& = 0.6404656, 0.6162334, 0.9579601 \nonumber \\ 
&c& = -0.0052439445, -0.0044137863, -0.0130500505 \nonumber
\end{eqnarray}
(for A2218, A2443, A119, respectively).  We use these analytical
representations of the `field' counts, adjacent to each cluster, for
statistical field subtraction throughout. We note that this method
will also correct, in a statistical sense, for any incompleteness in
the cosmic ray rejection.

Since we will use the analytic representation of the `field' counts
given in equation (\ref{eq:backcounts}) to remove contamination by `field' galaxies
superimposed on the cluster, we require an estimate of the uncertainty
in this analytic representation. To do this we use the formal variances
and covariances for each of the fit parameters ($a$, $b$ and $c$)
returned from the $\chi^2$ minimization.  The error ($\epsilon$) on the `field' counts
is given by:
\begin{eqnarray}
\label{eq:backerr}
\epsilon^2=\left({dN\over dV} \ln 10\right)^2 \left( \delta_a^2 + V^2\delta_b^2 + V^4 \delta_c^2 \right. \nonumber \\ \left. + 2V \delta_{ab} + 2V^2 \delta_{ac} + 2V^3\delta_{bc} \right)\end{eqnarray}
where $dN \over dV$ is the number of galaxies per magnitude interval
[given by equation (\ref{eq:backcounts})] centred on apparent magnitude $V$. $\delta_a^2$, $\delta_b^2$ and $\delta_c^2$ are the variances and $\delta_{ab}$, $\delta_{ac}$ 
and $\delta_{bc}$ are the covariances on the fitted parameters $a$, $b$ and $c$.

\subsection{Lensing}

A possible concern regarding the background counts is the impact of
lensing by the cluster mass. This consists of two effects: a
magnification effect whereby the background population is brightened,
and a tangential compression of the the background volume. While the
former can increase the counts, the latter can decrease it (i.e., the
effect is to narrow and lengthen the observable volume behind the
cluster). The precise details of this correction are given by
\citet{trenthamn98} and \citet{bernsteing95}.

In Fig.~\ref{fig:flens} we plot the fractional change in the observed number of
`field' galaxy counts caused by lensing, $f_{lens}$, as a function of
$V$-band magnitude for each cluster -- averaged over the central 500 kpc.
That is, the ratio of the number of `field' galaxies observed at a
given magnitude viewed along the cluster line-of-sight to the number
of galaxies expected if not observed through the cluster lens. In
modelling the lensing effect we follow the procedure outlined in
\citet{pracym04} and references therein.  We use a value for the
velocity dispersion of 1370\,km\,s$^{-1}$ for A2218
\citep{leborgnej92} and 778\,km\,s$^{-1}$ for A119
\citep{strublem87}; since no velocity data is available for A2443
we assume a velocity dispersion for this cluster of
1000\,km\,s$^{-1}$. We use the local luminosity function measured from
the MGC \citep{drivers05} and convert
to the $V$-band using $B-V=0.94$ \citep{norbergp02}.
\setcounter{figure}{7}
\begin{figure}
\includegraphics[height=8cm,angle=-90]{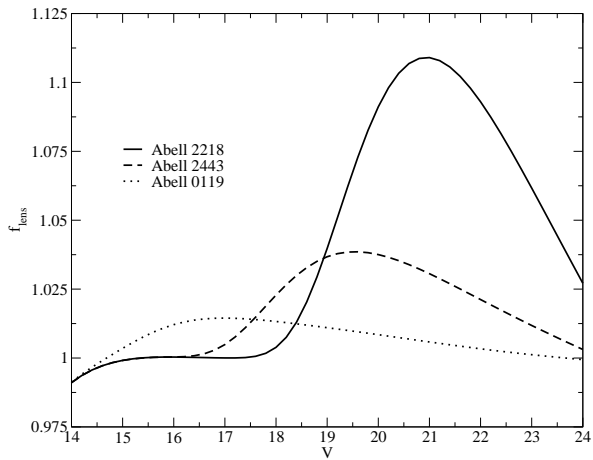}
\caption{\label{fig:flens}The function $f_{lens}$ averaged over the central 500 kpc for A2218 ({\it solid line}), A2443 ({\it dashed line}) 
and A119 ({\it dotted line}).} 
\end{figure}

In Fig.~\ref{fig:flens} the function $f_{lens}$ is shown for A2218 as the {\it solid
line}, for A2443 as the {\it dashed line} and for A119 as the {\it
dotted line}. Clearly, the effect is largest for A2218, primarily as a
result of the cluster's higher redshift. In A2218 the effect is most evident 
for galaxies with $V\approx 21$ with a fractional change in the
`background' population of approximately 10\%. At this level the
lensing effect represents a minor component of the error
budget. However it is worth noting that the lensing correction
depends critically on the assumptions concerning the evolution of the galaxy
number counts as a function of redshift, which is not well known.
Given this uncertainty and the small size of the corrections in
Fig.~\ref{fig:flens}, we do not correct for lensing at this time.

\section{Luminosity functions}
We now use our data along with our measured `field' counts
to study the luminosity distribution of galaxies in each cluster.  The
faint--end limits for A2218, A2443 and A119 are, respectively,
$M_V=-16.7$, $M_V=-15.4$ and $M_V=-13.3$\,mag.  These include
correction for galactic dust \citep{schlegeld98} and K-correction
\citep{poggiantib97}.

\subsection{Central luminosity distribution}
We first construct, for each cluster, the luminosity distribution of
galaxies within a cluster-centric radius of 1\,Mpc.  We use equation
(\ref{eq:backcounts}) -- scaled to the appropriate area -- to correct `statistically' for
contamination by the superimposed `field' galaxy population. These are
shown in the {\it left panels} of Fig.~\ref{fig:lf} for A2218 ({\it top}),
A2443 ({\it middle}) and A119 ({\it bottom}) -- and are also tabulated in Table \ref{tab:ld}. 
We fit each of the luminosity distributions with a \citet{schechterp76} function; which has been 
convolved with the magnitude bin width. The luminosity distributions in all three clusters 
are well described by a Schechter function with shape parameters
$M^*_{V}=-21.54_{-0.11}^{+0.14}$ and $\alpha=-1.14_{-0.07}^{+0.08}$
for A2218, $M^*_{V}=-21.32_{-0.36}^{+0.39}$ and
$\alpha=-1.11_{-0.09}^{+0.10}$ for A2443 and
$M^*_{V}=-21.51_{-0.20}^{+0.30}$ and $\alpha=-1.22_{-0.06}^{+0.07}$
for A119.  Since the parameters $M^*$ and $\alpha$ are dependent we plot, in the right hand panels of Fig.~\ref{fig:lf}, the
$1$, $2$ and $3\sigma$ error contours for the Schechter function fit.
\setcounter{figure}{8}
\begin{figure}
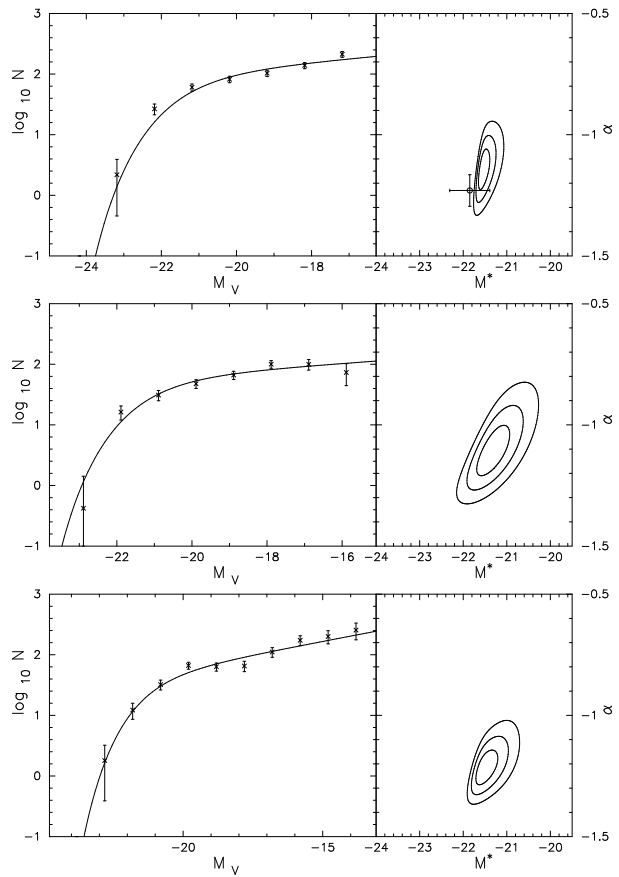

  \begin{center}
    \begin{minipage}{0.47\textwidth}
          \includegraphics[height=8.0cm, angle=-90, trim=0 0 0 0]{figure9a.ps}
          \includegraphics[height=8.0cm, angle=-90, trim=0 0 0 0]{figure9b.ps}
	  \includegraphics[height=8.0cm, angle=-90, trim=0 0 0 0]{figure9c.ps}
     \end{minipage}
    \caption{\label{fig:lf}Left panels: The luminosity distributions for the clusters A2218 ({\it top}), A2443 ({\it middle})
and A119 ({\it bottom}) within a cluster-centric radius of 1\,Mpc. The data are shown by the {\it stars} and 
the {\it solid line} is a \citet{schechterp76} function fit to the data. The right panels show the $1$, $2$ and $3\sigma$ 
error ellipses for the Schechter function parameters $M^*$ and $\alpha$. Also shown ({\it open circle in top panel}) is the best--fitting Schechter function
parameters for the `bright-end' of the  luminosity distribution of A2218 from \citet{pracym04}.}
    \end{center}
\end{figure}
\begin{table}
\setcounter{table}{1}
\caption{\label{tab:ld}luminosity distributions}
\centering
\begin{tabular}{|c|c|c|c|} \hline
Cluster  &  Mag interval & Counts        &   Error      \\ \hline
A2218    &  $-23.68 < M_{V} \le -22.68$  &   2.19      &  1.73        \\ 
 ''      &  $-22.68 < M_{V} \le -21.68$  &   26.67     &  5.39        \\
 ''      &  $-21.68 < M_{V} \le -20.68$  &   60.48     &  8.19        \\
 ''      &  $-20.68 < M_{V} \le -19.68$  &   82.16     &  10.00       \\
 ''      &  $-19.68 < M_{V} \le -18.68$  &   103.37    &  12.29       \\
 ''      &  $-18.68 < M_{V} \le -17.68$  &   138.87    &  16.22       \\
 ''      &  $-17.68 < M_{V} \le -16.68$  &   212.23    &  22.98       \\ 
\\
 A2443   &  $-23.39 < M_{V} \le -22.39$  &   0.42      &  1.00        \\ 
 ''      &  $-22.39 < M_{V} \le -21.39$  &   16.28     &  4.24        \\
 ''      &  $-21.39 < M_{V} \le -20.39$  &   31.01     &  6.00        \\ 
 ''      &  $-20.39 < M_{V} \le -19.39$  &   47.78     &  7.87        \\
 ''      &  $-19.39 < M_{V} \le -18.39$  &   66.31     &  10.30       \\ 
 ''      &  $-18.39 < M_{V} \le -17.39$  &   100.41    &  14.46       \\
 ''      &  $-17.39 < M_{V} \le -16.39$  &   99.91     &  19.78       \\ 
 ''      &  $-16.39 < M_{V} \le -15.39$  &   73.38     &  28.96       \\
\\
 A119    &  $-23.31 < M_{V} \le -22.31$  &   1.80      &  1.41        \\ 
 ''      &  $-22.31 < M_{V} \le -21.31$  &   12.24     &  3.61        \\
 ''      &  $-21.31 < M_{V} \le -20.31$  &   32.24     &  5.92        \\ 
 ''      &  $-20.31 < M_{V} \le -19.31$  &   66.54     &  8.72        \\
 ''      &  $-19.31 < M_{V} \le -18.31$  &   63.48     &  9.72        \\ 
 ''      &  $-18.31 < M_{V} \le -17.31$  &   65.33     &  12.72       \\
 ''      &  $-17.31 < M_{V} \le -16.31$  &   111.00    &  20.16       \\ 
 ''      &  $-16.31 < M_{V} \le -15.31$  &   173.38    &  32.21       \\
 ''      &  $-15.31 < M_{V} \le -14.31$  &   200.41    &  49.05       \\ 
 ''      &  $-14.31 < M_{V} \le -13.31$  &   254.88    &  78.41       \\ \hline
\end{tabular}
\end{table}

The luminosity distribution of galaxies in A2218 has been studied by \citet{pracym04} using a deep HST mosaic of the cluster. Their `bright-end' luminosity distribution is comparable in bandpass, areal coverage and luminosity range to the one derived above -- with a projected area covered of $\sim 2.6\, h^{-2}_{0.7}$\,Mpc$^2$ (compared with $\sim 3.1\, h^{-2}_{0.7}$\,Mpc$^2$ in this study) and a luminosity range corresponding to $M_{V}<-17$ (compared with $M_{V}<-16.7$ here). Here we have converted to the $V$-band using the relation $V=F606W+0.12$ \citep{norbergp02,drivers03} and corrected for differences in assumed cosmology, dust extinction, K-correction, and the omission of the cD galaxy in \citet{pracym04}. We show the best--fitting Schechter function parameters from \citet{pracym04} along with the error contours in Fig.~\ref{fig:lf}. The recovered Schechter function parameters are in good agreement ($-1.23\pm 0.09$ cf. $\alpha=-1.14_{-0.07}^{+0.08}$ in this study and $M^*_{V}=-21.85\pm 0.46$ compared with $M^*_{V}=-21.54_{-0.11}^{+0.14}$ here).

\begin{table*}
\setcounter{table}{2}
\caption{\label{tab:litcompare}Recent measurements of the LF parameters from wide--field $V$-band CCD photometry}
\begin{tabular}{|l|c|c|c|c|c|l|} \hline
Cluster     & redshift & Area (Mpc$^2$)      & Magnitude range           &   $M_{V}^{*}$               & $\alpha$                 & Reference            \\  \hline
Coma$\dagger$        &  0.023   & $\sim 0.9$ &     $-21.51<M_{V}<-11.51$ &   NA                        & $-1.43_{-0.03}^{+0.06}$    & \citet{andreons02b}  \\
Coma        &  0.023   & $\sim 1.1$ &     $-21.51<M_{V}<-14.01$ &   NA                        & $-1.59\pm0.02$           & \citet{loboc97}      \\
Abell 2151  &  0.037   & $\sim 8.3$ &     $M_{V}< -15.0$        & $-21.56_{-0.41}^{+0.44}$    & $-1.29_{-0.08}^{+0.09}$  & \citet{sanchezr05}   \\
Abell 119  &  0.044   & $\sim 3.1$ &     $M_{V}< -13.3$        & $-21.51_{-0.20}^{+0.30}$    & $-1.22_{-0.06}^{+0.07}$  & This work            \\
Abell 2443  &  0.108   & $\sim 3.1$ &     $M_{V}< -15.4$        & $-21.32_{-0.36}^{+0.39}$    & $-1.11_{-0.09}^{+0.10}$  & This work            \\
Abell 2218  &  0.181   & $\sim 3.1$ &     $M_{V}< -16.7$        & $-21.54_{-0.11}^{+0.14}$    & $-1.14_{-0.07}^{+0.08}$  & This work            \\
ABCG 209    &  0.209   & $\sim 3.4$ &     $M_{V}< -17.6$        & $-22.18\pm 0.3$             & $-1.27\pm 0.10$          & \citet{mercurioa03}  \\
ABCG 209    &  0.209   & $\sim 6.9$ &     $M_{V}< -17.6$        & $-22.03\pm 0.3$             & $-1.25\pm 0.08$          & \citet{mercurioa03}  \\
AC 118      &  0.310   & $\sim 5.5$ &     $M_{V}< -18.2$        & $-20.93\pm 0.4$             & $-1.02\pm 0.18$          & \citet{busarellog02} \\ \hline
\end{tabular}

\noindent $\dagger$ Based on a Schechter function fit to on-line data \\
\end{table*}

In Table \ref{tab:litcompare} we tabulate the results of  recent wide--field $V$-band LF studies: column (1) identifies the cluster, column (2) gives the cluster redshift, column (3) gives the total area covered, column (4) indicates the absolute magnitude limits of the data, column (5) presents the characteristic magnitude, column (6) gives the value of the faint--end--slope parameter, and column (7) identifies the reference. All values have been converted to an ${\Omega}_{M}=0.3$, ${\Omega}_{\Lambda} =0.7$ and $H_{0}=70$\,km\,s$^{-1}$\,Mpc$^{-1}$
cosmology. Although, comparison of the LF parameters is somewhat complicated by differences in redshift (and hence rest-frame bandpass), areal coverage, and the magnitude limits of the data -- the values, and the precision, of the LF parameters presented here are comparable to those given in the literature. However, we derive faint--end--slope parameters for these clusters which are marginally flatter than the average. With the exception of \citetalias{andreons02b} Coma LF, our Abell 119 data represents the deepest determination of a wide--field $V$-band LF available.

\subsection{Composite luminosity distribution}
We now construct a composite luminosity distribution from our three clusters. To
do this we average the background--subtracted number counts in each magnitude bin normalised by the
cluster's `richness': 
\begin{equation}
\label{eq:lfcomp}
\zeta(M)={\sum_{i=1}^{n_{\rm c}} N_i(M) \over \sum_{i=1}^{n_{\rm c}} R_i},
\end{equation}
where   $\zeta(M)$ is the 
combined galaxy number `counts' in the magnitude interval centred on $M$, $n_{\rm c}$ 
is the number of clusters in the composite luminosity distribution ($n_{\rm c}=3$) and 
$N_{i}(M)$ is the galaxy number counts in the magnitude bin centred on $M$ for the $i$th 
cluster.  $R_{i}$ is the `richness count' of the $i$th cluster, which is taken to be 
the total number of background--subtracted galaxies in the magnitude range $-21<M_{V}<-19$ within 
a cluster-centric radius of $1.5\; h_{0.7}^{-1}$\,Mpc \citep{valottoc01}.

The composite luminosity distribution and best--fitting Schechter function ({\it solid line})
is shown in Fig.~\ref{fig:lfcomb}. The shape of the best--fitting Schechter function is parameterised by  
$M^*_{V}=-21.82_{-0.18}^{+0.21}$ and $\alpha=-1.22_{-0.04}^{+0.05}$.
The $1$, $2$ and $3\sigma$ error contours of the parameters $M^*_{V}$ and $\alpha$ for our 
composite luminosity distribution are shown in the {\it right panel} of Fig.~\ref{fig:lfcomb}, along 
with the positions in the $M^*_{V}$--$\alpha$ plane of a subsample of the composite luminosity
function parameters from the literature (converted to the $V$-band).  
The {\it open circle} and {\it star} represent the 
field galaxy luminosity functions derived from the MGC \citep{drivers05} and Two Degree Field Galaxy Redshift
Survey \citep[2dFGRS; ][]{collessm01}, both of which exhibit faint--end slopes ($\alpha$) that are similar to that of our composite luminosity 
function but have a brighter and fainter characteristic magnitude ($M^*$), respectively. The {\it open square} represents the best--fitting
Schechter function parameters for a composite of clusters \citep{deproprisr03} drawn from the 2dFGRS
with a mean redshift of $z=0.12$. Our composite luminosity function
has a marginally brighter characteristic magnitude to that of the 2dFGRS cluster composite and a slightly flatter $\alpha$. The {\it open triangle} 
shows the formal Schechter function parameters from \citet{gotot02} for a composite of clusters in the redshift range 
$0.2<z<0.25$ from the Sloan Digital Sky Survey, which has a significantly brighter characteristic magnitude and flatter
faint--end slope.    
\setcounter{figure}{9}
\begin{figure}
\includegraphics[height=8cm,angle=-90]{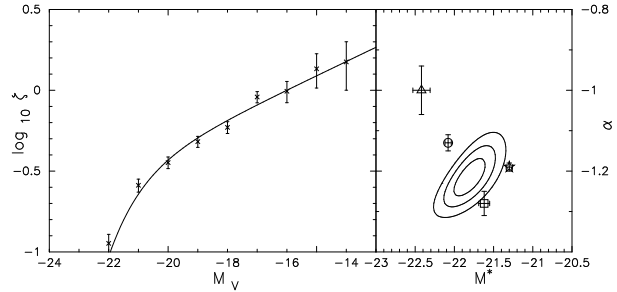}
\caption{\label{fig:lfcomb}Left panel: The composite luminosity function of the three clusters.  Right panel: the $1$, $2$
and $3\sigma$ error ellipses with a set of Schechter function parameters for composite luminosity functions from the literature
superimposed. See text for details. }
\end{figure}

\section{Geometric deprojection}
The wide field-of-view of the WFC mosaics provides coverage
of the entire cluster area, in the sense that the number of galaxies
per unit area no longer decreases with radius at the outermost regions
covered by the imaging.

We use the scheme of \citet[][see
\citealt{beijersbergenm02}]{fabiana81} to perform geometric
deprojection of the clusters.  The projected cluster galaxy population
is separated into a set of concentric annuli within which the galaxy
density is assumed to be constant.  The deprojection scheme assumes
the cluster galaxies are distributed spherically symmetrically so that
the projected number of galaxies in each annulus corresponds to a real
galaxy number density in the corresponding spherical shell.  The
number density in each shell can then be corrected for the projection
of galaxies from all shells of greater radius, beginning with the
outermost shell and iteratively moving inward.

 The projected galaxy density for each magnitude bin, $m$, in the $i$th annulus, $N_{i,m}$, is given by:
\begin{eqnarray}
\label{eq:depro}
N_{i,m}=\phi_{i,m}F(R_{i},R_{i-1},R_{i})+ \nonumber \\ \sum_{j=i+1}^n\phi_{j,m}(F(R_{j},R_{i-1},R_{i})-F(R_{j-1},R_{i-1},R_{i}))
\end{eqnarray}
where the $R_{i}$'s are the radii of the annuli, $\phi_{i,m}$ is the 3D galaxy density and the function $F$ is defined by:
\begin{eqnarray}
\label{eq:geom}
F(\alpha,\beta,\gamma)={4\over 3}\pi\alpha^3\biggl(\biggl(1-{\beta^2 \over \alpha^2}\biggl)^{3/2}-\biggl(1-{\gamma^2 \over \alpha^2}\biggl)^{3/2}\biggl)
\end{eqnarray}
 \citep{beijersbergenm02}.

We set radial bin partitions at cluster-centric radii of 0.3, 0.6 and 1.5\,Mpc -- splitting the cluster field into four regions ($n=4$) --  and perform geometric deprojection of the data in 1 magnitude intervals.  We first subtract from the number counts in each annuli the contribution from the superimposed field galaxy population using equation (\ref{eq:backcounts}) -- normalised to the area of `detectability'  for each annulus. We use the outermost annuli ($R>1.5$\,Mpc) to begin the deprojection, that is, set $i=3$ in equation (\ref{eq:depro}) and then iteratively calculate the deprojected (3-Dimensional) luminosity function in each of the inner three annuli: $0.0\,{\rm Mpc}<r<0.3\,{\rm Mpc}$, $0.3\,{\rm Mpc}<r<0.6\,{\rm Mpc}$ and $0.6\,{\rm Mpc}<r<1.5\,{\rm Mpc}$.  The errors in the 3D galaxy density are calculated by propagating the appropriate galaxy--count and field--subtraction errors through equation (\ref{eq:depro}).

The deprojected 3D luminosity functions are shown in Fig.~\ref{fig:depro} for A2218 ({\it top}), A2443 ({\it middle}) and A119 ({\it bottom}). The {\it open circles, open squares} and {\it open triangles} are the galaxy densities in the inner, middle and outer annuli, respectively, and the {\it solid, dashed} and {\it dotted lines} show their best fitting Schechter functions. The $1$, $2$ and $3\sigma$ error contours for the Schechter function shape parameters $M_{V}^*$ and $\alpha$ are shown in the right hand panels of Fig.~\ref{fig:depro}. The core region ($r<300$ kpc) of A2218 has a LF with a  flatter faint--end slope than the LFs measured at larger cluster-centric radii within this cluster, while the other clusters show no significant systematic correlation with cluster-centric radius.  A summary of the Schechter function fit parameters are given in Table \ref{tab:depro}.
\begin{table}
\setcounter{table}{3}
\caption{\label{tab:depro}Schechter function parameter values for the deprojected (3-dimensional) luminosity functions
in three annuli: $0.0\,{\rm Mpc}<r<0.3\,{\rm Mpc}$, $0.3\,{\rm Mpc}<r<0.6\,{\rm Mpc}$ and $0.6\,{\rm Mpc}<r<1.5\,{\rm Mpc}$}
\centering
\begin{tabular}{|l|l|l|l|} \hline
Cluster  & Radius (Mpc) & $M_{V}^*$ & $\alpha$                \\ \hline
A2218    &  0--0.3       & $-22.12_{-0.57}^{+0.57}$ & $-0.94_{-0.16}^{+0.19}$   \\ 
         &  0.3--0.6     & $-22.31_{-0.77}^{+0.60}$ & $-1.30_{-0.15}^{+0.16}$   \\
         &  0.6--1.5     & $-21.86_{-0.35}^{+0.37}$ & $-1.36_{-0.10}^{+0.11}$   \\

A2443    &  0--0.3       & $-20.72_{-0.63}^{+0.58}$ & $-1.00_{-0.19}^{+0.22}$   \\
         &  0.3--0.6     & $-21.72_{-0.63}^{+0.72}$ & $-1.19_{-0.12}^{+0.14}$   \\
         &  0.6--1.5     & $-20.03_{-0.88}^{+1.02}$ & $-1.01_{-0.18}^{+0.45}$   \\ 

A119    &  0--0.3       & $-22.24_{-1.78}^{+1.54}$ & $-1.25_{-0.11}^{+0.16}$   \\ 
         &  0.3--0.6     & $-21.53_{-0.70}^{+0.71}$ & $-1.16_{-0.12}^{+0.16}$  \\
         &  0.6--1.5     & $-21.66_{-0.54}^{+0.51}$ & $-1.23_{-0.10}^{+0.12}$   \\ \hline
\end{tabular}
\end{table}
\setcounter{figure}{10}
\begin{figure}
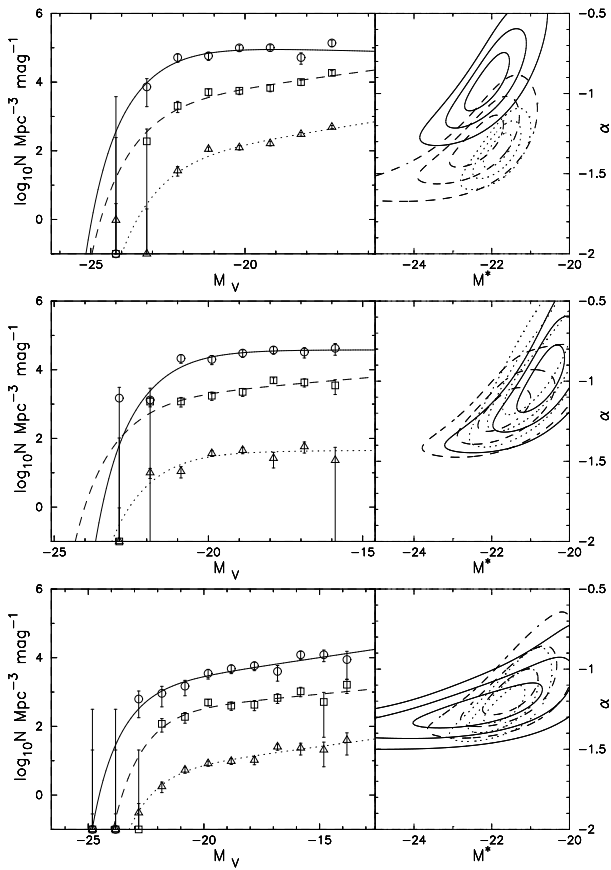

  \begin{center}
    \begin{minipage}{0.47\textwidth}
          \includegraphics[height=8.0cm, angle=-90, trim=0 0 0 0]{figure11a.ps}
          \includegraphics[height=8.0cm, angle=-90, trim=0 0 0 0]{figure11b.ps}
	  \includegraphics[height=8.0cm, angle=-90, trim=0 0 0 0]{figure11c.ps}
     \end{minipage}
    \caption{\label{fig:depro}Left panels: the deprojected luminosity distributions of galaxies in cluster-centric annuli of 0--0.3\,Mpc ({\it open circles}),
0.3--0.6\,Mpc ({\it open squares}) and 0.6--1.5\,Mpc ({\it open triangles}) and their best fitting Schechter functions ({\it solid, dashed and dotted lines}, respectively). Right panels: $1$, $2$ and $3\sigma$ error contours for the Schechter function parameters $M_{V}^*$ and $\alpha$.  A2218: top panel, 
A2443: middle panel and A119: bottom panel.}
    \end{center}
\end{figure}

\section{Summary and Discussion}
We have exploited wide--field imaging of three galaxy clusters to examine the luminosity and spatial distribution of their member galaxies and to search for evidence of luminosity segregation.  We cover a total projected area which corresponds to 120.5\,Mpc$^2$, 50.4\,Mpc$^2$ and 9.7\,Mpc$^2$ in A2218, A2443 and A119, respectively.  We find that the radial distribution of galaxies can be well described by a King profile with a core radius which is essentially independent of luminosity -- the exception being the brightest galaxies in A2218 which exhibit a more compact spatial distribution.  Galaxies at different luminosities having the same spatial distribution, of course, indicates that no luminosity segregation is present. We find, in all three clusters, a luminosity distribution of galaxies which is well described by a Schechter function with a flat faint--end slope ($\alpha\sim -1.2$). We have performed a geometric deprojection of the cluster population. The luminosity function remains flat after deprojection and (except for the core region of A2218) we find no statistically significant evidence for any change with cluster-centric radius -- consistent with our radial profile analysis.  

The segregation of galaxies of different luminosities is a key prediction of hierarchical models of galaxy formation and evolution. The numerical CDM models of \citet{kauffmanng97}, for example, predict that low luminosity galaxies should be less clustered than the bright `giant' galaxies \citep[see][]{phillippss98}. Evolutionary mechanisms operating in clusters which should give rise to luminosity segregation have also been suggested.  \citet{mooreb98} have been able to explain the dwarf population--density relation (in which the ratio of dwarf galaxies to giant galaxies increases with decreasing local galaxy density) as the result of galaxy `harassment' -- which operates more effectively in regions of high galaxy density.  Interaction with the cluster tidal field will destroy the lowest surface brightness (least bound) objects in the central region of a cluster, resulting in a deficiency of faint galaxies in the cluster core.  If luminosity segregation in clusters is not present then the impact of these effects on the low-luminosity cluster galaxies has been exaggerated.    

The key issue in searching for luminosity segregation in clusters is that of the technique of background subtraction employed. The fractional difference between the total (cluster+`field') galaxy counts and the `field' galaxy counts decreases with decreasing luminosity. Therefore, any systematic error in the estimate of the field galaxy counts can mimic both a faint--end upturn in the luminosity function and luminosity segregation -- in the sense that underestimating the field counts will result in a larger fraction of faint galaxies at low local galaxy densities. Using spectroscopy to confirm cluster membership is essentially useless for all but the nearest clusters, since the galaxies at the critical faint end of the luminosity function are too faint for spectroscopic confirmation. Hence, most often a statistical  subtraction of the field is performed using galaxy counts measured from a region offset from the cluster line-of-sight. \citet{valottoc01} point out one possible source of bias in statistical field subtraction due to the alignments between filaments and clusters which can result in an underestimate of the `field' galaxy population. Here we have used an analytic representation (equation \ref{eq:king}) of the radial distribution of galaxies toward the cluster sight-line to simultaneously obtain the cluster profile and the superimposed `field' galaxy surface density.  Note, fitting equation (\ref{eq:king}) to the radial profiles in Fig.~\ref{fig:rad1}, Fig.~\ref{fig:rad2} and Fig.~\ref{fig:rad3} does not require prior subtraction of the field galaxy population and we, therefore, suggest that this is a superior test for luminosity segregation.  Very recently \citet{andreons05} have suggested an alternative method to alleviate and quantify the problems of background subtraction in deriving cluster luminosity functions -- a detailed comparison of these methods and the traditional method should be pursued in a future paper.

\def\ref{\par\noindent\hangindent\parindent}

\section*{Acknowledgements}
Based on observations made with the Isaac Newton Telescope operated on
the island of La Palma by the Isaac Newton Group in the Spanish
Observatorio del Roque de los Muchachos of the Instituto de
Astrofisica de Canarias. We thank the Cambridge Astronomical Survey
Unit for reducing the INT data. 
We also thank the anonymous referee for a very helpful report which has greatly 
improved this paper and Chris Blake for his help and advice with this work.
M.B.P. was supported by an Australian Postgraduate Award.  S.P.D and W.J.C. acknowledge the financial
support of the Australian Research Council throughout the course of
this work. This research has made use of the NASA/IPAC Extragalactic
Database (NED) which is operated by the Jet Propulsion Laboratory,
California Institute of Technology, under contract with the National
Aeronautics and Space Administration.

\bibliographystyle{mn2e}
\bibliography{references}

\clearpage

\label{lastpage}

\end{document}